\documentclass[12pt]{article}
\usepackage{epsfig}
\usepackage{a4,isolatin1}
\usepackage{amsmath,amsfonts,latexsym, amssymb}

\newtheorem{satz}{Theorem}[section]
\newtheorem{defi}[satz]{Definition}

\newtheorem{bem}[satz]{Remark}

\newtheorem{koro}[satz]{Corollary}
\newtheorem{bsp}[satz]{Example}

\newtheorem{obdef}[satz]{Observation/Definition}
\newtheorem{conclusion}[satz]{Conclusion}
\newtheorem{ob}[satz]{Observation}

\newtheorem{stat}[satz]{The Statistical Hypothesis}
\newtheorem{random}[satz]{The Random Graph Idea}
\newtheorem{variant}[satz]{A Variant Approach}
\newtheorem{res}[satz]{R\'esum\'e}
\newtheorem{points}[satz]{The Qualitative Picture}

\newcommand{\mcal}{\mathcal}

\newcommand{\tit}{\textit}

\newcommand{\R}{\mathbb{R}}
\newcommand{\Z}{\mathbb{Z}}

\begin{document}
\thispagestyle{empty}
\begin{center}
\vspace*{1.0cm}

{\LARGE{\bf (Quantum) Space-Time as a\\ Statistical Geometry of
    Lumps\\ in Random Networks}}

\vskip 1.5cm

{\large {\bf Manfred Requardt}}\\email: requardt@theorie.physik.uni-goettingen.de 

\vskip 0.5 cm 

Institut f\"ur Theoretische Physik \\ 
Universit\"at G\"ottingen \\ 
Bunsenstrasse 9 \\ 
37073 G\"ottingen \quad Germany

\end{center}

\vspace{1 cm}

\begin{abstract}
  In the following we undertake to describe how macroscopic space-time
  (or rather, a microscopic protoform of it) is supposed to emerge as a
  superstructure of a web of lumps in a stochastic discrete network
  structure. As in preceding work (mentioned below), our analysis is
  based on the working philosophy that both physics and the
  corresponding mathematics have to be genuinely discrete on the
  primordial (Planck scale) level. This strategy is concretely
  implemented in the form of \tit{cellular networks} and \tit{random
    graphs}. One of our main themes is the development of the concept
  of \tit{physical (proto)points} or \tit{lumps} as densely entangled
  subcomplexes of the network and their respective web, establishing
  something like \tit{(proto)causality}. It may perhaps be said that
  certain parts of our programme are realisations of some early ideas
  of Menger and more recent ones sketched by Smolin a couple of years
  ago. We briefly indicate how this \tit{two-story-concept} of
  \tit{quantum} space-time can be used to encode the (at least in our
view) existing non-local aspects of quantum theory without violating
macroscopic space-time causality!

\end{abstract} \newpage
\setcounter{page}{1}
\section{Introduction}
In recent papers (\cite{1} to \cite{4}) we developed some facets of an
extensive programme we formulated there, i.e. reconstruct ordinary
continuum physics or mathematics as kind of a coarse grained limit from
a much more primordial and genuinely discrete substrate, including, in particular, a discrete theory of
(proto) space-time as the universal receptacle or substratum of all
physical processes.

A corresponding philosophy is presently hold by a substantial minority
of workers in \tit{quantum gravity} and we commented on some of the
various approaches, at least as far as we are aware of them, in the
foregoing papers. We therefore refer the interested reader to these
papers as to references we do not mention in the following just for
sake of brevity. As an exception we mention only the early and
prophetic remarks made by Penrose in e.g. \cite{5} about the surmised
combinatorial substratum underlying our continuous space-time , the
ideas of Smolin, sketched at the end of \cite{6}, because they are
surprisingly close to our working philosophy and the work of 't Hooft
(\cite{7}) which is based on the model system of \tit{cellular
  automata} (the more rigid and regular relatives of our \tit{dynamic
  cellular networks} introduced in the following).

Our personal working philosophy is that space-time at the very bottom
(i.e. near or below the notorious Planck scale) resembles or can be
modeled as an evolving information processing cellular network,
consisting of elementary modules (with, typically, simple internal
discrete state spaces) interacting with each other via dynamical bonds
which transfer the elementary pieces of information among the nodes.
That is, the approach shares the combinatorial point of view in
foundational space-time physics which has been initiated by e.g.
Penrose. It is a crucial and perhaps characteristic extra ingredient
of our framework that the bonds (i.e. the elementary interactions) are
not simply dynamical degrees of freedom (as with the nodes their
internal state space is assumed to be simple) but can a fortiori,
depending on the state of the local network environment, be switched
on or off, i.e. can temporarily be active or inactive! This special
ingredient of the dynamics hopefully allows the network to perform
\tit{geometric phase transitions} into a new ordered phase having a
certain \tit{two-story structure} to be explained below. This
conjectured geometric order we view as kind of a discrete \tit{proto
  space-time} carrying metrical, causal and dimensional
structures. The main content of the paper consists of a description
and analysis of these structural elements and their interplay. 

In paper \cite{1} we dealt primarily with dimensional
concepts on such discrete and irregular spaces. It furthermore became
aparent that there exist close ties to the theory of \tit{fractal
  sets}. Papers \cite{3} and \cite{4}, on the other hand, are, among
other things, devoted to the developement of several possible versions of
\tit{discrete analysis} and \tit{discrete differential geometry}
respectively \tit{discrete functional analysis} with certain zones of
contact with \tit{noncommutative geometry}.

As has been beautifully reviewed by Isham in various papers (see e.g.
\cite{8}) one could, among several possible attitudes, adopt the
perhaps most radical working philosophy in quantum gravity and
speculate that both \tit{quantum theory} and \tit{gravity} are merely
secondary and derived aspects or, expressed in more physical terms,
so-called \tit{effective theories} of an underlying more primordial
theory of a markedly \tit{combinatorial} flavor. But a theory
comprising quantum theory and gravitation as \tit{emergent}
subtheories should, first of all, provide a framework in which both
the emergence of something we may consider as a proto form of
classical space-time respectively \tit{quantum vacuum} can be
expressed or discussed, most notably the emergence of the
\tit{continuum} from the \tit{discrete} and the concept of physical
\tit{space-time points} (having an internal dynamical structure) and
their intricate web. In a companion paper, which is in preparation, we
will show how \tit{quantum theory} emerges as a coarse grained
stochastic theory within such a framework.

Some months ago, after the preparation of the first draft, we were
kindly informed by El Naschie that a very early source where a couple
of related ideas can be found is the contribution of Menger in
\cite{Me}, perhaps better known from his research on topological
dimension or fractal sets like the Menger sponge. In this essay he
entertains very interesting ideas about the necessity of a new
\tit{geometry of the microcosmos} based on the \tit{geometry of lumps}
and the concept of a \tit{statistical metric space}. We note in
passing that some of the concepts we develop in the following seem to
be very much in accord with this point of view.  Quite remarkable in
this respect is also Einsteins openminded attitude towards the
possibility of a more primordial discrete space-time theory expressed
at the end of the same volume. As to this particularly important point
see also his many utterances compiled by Stachel in \cite{Einstein},
p.27ff. We think, our discrete framework may exactly realize some of
the ideas he had in mind. (In this context we also want to mention the
\tit{Cantorian space-time approach} of El Naschie et al. who tries to
model microspace as a particular type of \tit{(random) fractal} (see
e.g.  \cite{Naschie}).  Another interesting early source with a
possible bearing on our approach may also be v.Neumann´s concept of
\tit{continuum geometry} (geometry without points) which is briefly
described in \cite{Neumann}, the original source can be found in
\cite{NeumannColl}. Furthermore, there exist surprising links to other
seemingly unrelated areas of current research as we learnt very
recently, a catchword being \tit{small-world networks} (see
\cite{10}). Some of the mechanisms being effective in our cellular
network environment are also observed there.

After having nearly completed this second version we received a
message from S.Roy who informed us that the framework sketched by
Menger has been developed further by Menger himself together with
several coworkers and later by other groups (seee e.g. \cite{Rosen}).
The state of affairs is described in his monograph (\cite{Roy}), the
more mathematical aspects are developed in \cite{Schweizer}. From
these monograhs one may see that the early ideas of Menger have
meanwhile unfolded into various branches of their own right and it is
not easy at the moment to relate all of them with our own working
philosophy. Anyhow, the situation looks very promising and
interesting; we will deal with the geometric web of lumps (we also
called them \tit{cliques}) in section 4 (which is the central section
of the paper) and make some remarks concerning the \tit{metric}
aspects in subsection 4.3.

Some clarifying comments are perhaps in order at this point. The
modelling of the depth structure of space-time as a cellular network
consisting of nodes and bonds should not necessarily be understood in
a plain bodily sense. One should rather consider it as a
representation or emulation of the main characteristics of the
physical scenario. In this connection we want to mention the following
interesting and purely discrete approaches in  \cite{Sorkin} and \cite{Kauffman}.There may, in particular, exist a variety of
superficially different systems, the logical structure of which can
nevertheless be encoded in roughly the same abstract underlying
network model. It is our belief that such a discrete network,
governed by a relatively simple but cleverly chosen dynamical law, is
capable of generating most if not all of the phenomena and emergent
laws on which our ordinary continuum physics is usually grounded. That
such a hypothesis is not entirely far-fetched may e.g. be inferred
from the emerging complexity of such a simple cellular automaton model
as the famous \tit{game of life} created by Conway. A typical example
is the \tit{geometry of lumps} envisaged by Menger. Take as lumps the
hypothetical \tit{infinitesimal} grains of space or space-time which
cannot be further resolved (be it in a practical or principal sense).
Let them overlap according to a certain rule so that they can interact
or exchange information. Draw a node for each such lump and a bond for
each two lumps which happen to overlap. In \tit{combinatorial
  topology} such a \tit{combinatorial complex} is called the
\tit{nerve} of the \tit{set system} (cf. e.g.  \cite{Hopf}). In a next
step one may encode the respective strengths of interaction or degrees
of overlap in a valuation of the corresponding bonds, yielding a
cellular network of the kind we are having in mind. A fortiori one can
make these mutual overlaps into dynamical variables, i.e. let them
change in the course of evolution.

It may seem, at first glance, that our \tit{bottom-up approach}, which
starts almost from first principles, has the disadvantage, if compared
with other more \tit{top-down} oriented approaches as e.g. \tit{string
  theory} or \tit{loop quantum gravity}, to be rather arbitrary. These
other frameworks are based (at least to a large part) on more or less
continuum concepts and ideas theoretical physics has got accustomed to
and detect some \tit{discrete behavior} (e.g. \tit{spin networks})
only at the end of the road if at all. Furthermore, in the case of
e.g. string theory, the \tit{quantum principles} (whatever that means
exactly) are taken for granted down to arbitrarily fine scales (at
least as far as we can see) and both rely on such approved concepts as
\tit{functional integrals} or \tit{sum over configurations}. To such a
potential criticism we would like to reply in the following way. For
one, it is far from obvious that the (typically euclidean) functional
integral philosophy does still hold sway in e.g. the Planck regime, as
it is (implicitly) based on some kind of \tit{action principle} which,
on its side, may have its roots in classical physics. On the other
side, it may turn out that such concepts are only the surface aspect
of a deeper, more hidden reality. Such a possibility can of course
not be precluded, but we do not want to base our approach on such a
heuristic principle from the outset. Quite to the contrary, we want to
deduce quantum theory (and by the same token the functional integral
framework) as an \tit{effective theory} from our presumably more
fundamental theory. How this may work we undertake to show in our above
mentioned companion paper. We expect however that these approaches may
merge on some intermediate level.
 
We would like to add yet another, as we think, important remark. Quite
a few researchers in this field expect physics at the Planck scale to
be quite different from the kind of physics we are accustomed to (take
e.g. the interesting remarks in \cite{Nielsen}). The situation may be
comparable to the physics near or at the \tit{critical point} in, say,
\tit{renormalisation theory} where different microscopic theories (the
so-called \tit{universality class}) lead to basically the same
macroscopic behavior as the microscopic details happen to be washed
out in the coarse-graining process. This does of course not mean that
the search for such a underlying microscopic theory is futile or
useless. The lesson is rather that one should be prepared to employ
perhaps other guiding principles in the quest for this presumed more
primordial laws and concentrate, at least in a first step, rather on a
whole class of possible primordial model theories instead of a single
\tit{theory of everything}. As we will show in the mentioned companion
paper, quantum behavior may just emerge as such a coarse-grained
stochastic behavior of a whole class of discrete microscopic model
theories. In other words, one of our more heuristic guiding principles
is the following. We consider a class of microscopic (discrete)
theories worth to be studied if they have the propensity to generate
e.g. \tit{quantum behavior} or a kind of \tit{proto space-time} in
some classical limit. It would even be nicer if, at an intermediate
level, they develop links to, say, \tit{string theory} or \tit{loop
  quantum gravity}. That there are relations to \tit{non-commutative
  geometry} was already shown in the above mentioned papers.

\section{The Cellular Network Environment}
We briefly describe in this section the general framework within which
our investigation will take place. Among other things we introduce a
simple example of a local dynamical network law (for more details see
\cite{Planck} or \cite{3}. A certain class of relatively simple
  cellular networks is the following.
\begin{defi}[Class of Cellular Networks] \hfill  
\begin{enumerate}
\item ``Geometrically'' our networks represent at each fixed
{\em `clock time'} {\em `labeled graphs'}, i.e. they consist of nodes
\{$n_i$\} and bonds \{$b_{ik}$\}, with the bond $b_{ik}$ connecting
the nodes (cells) $n_i$, $n_k$. We assume that the graph has neither
elementary loops nor multi-bonds, that is, only nodes with $i\neq k$
are connected by at most one bond.
\item At each site $n_i$ we have a local node state $s_i\in
q\cdot\Z$ with $q$, for the time being, a certain not further
specified  elementary quantum. The bond variables $J_{ik}$, attached
to $b_{ik}$, are in the most simplest cases assumed to be two- or
three-valued, i.e. $J_{ik}\in \{\pm 1\}\quad\mbox{or}\quad J_{ik}\in \{\pm
  1,0\}$
\item There exist a variety of possibilities concerning the class of
  bonds being active in the network. The simplest and perhaps most
  natural choice is to assume that, at least in the \tit{initial
    configuration} (see below), every pair of nodes is connected by a
  bond. On the other hand, one can restrict the set of admissible
  bonds to a certain subset (motivated perhaps by the physical context).
\end{enumerate}
\end{defi}

A simple example of such a local dynamical law we are having in mind
is given in the following definition.
\begin{defi}[Example of a Local Law]
At each clock time step a certain `{\em quantum}' $q$
is exchanged between, say, the nodes $n_i$, $n_k$, connected by the
bond $b_{ik}$ such that 
\begin{equation} s_i(t+\tau)-s_i(t)=q\cdot\sum_k
  J_{ki}(t)\end{equation}
(i.e. if $J_{ki}=+1 $ a quantum $q$ flows from $n_k$ to $n_i$ etc.)\\
The second part of the law describes the {\em back reaction} on the bonds
(and is, typically, more subtle). This is the place where the
so-called `{\em hysteresis interval}' enters the stage. We assume the
existence of two `{\em critical parameters}'
$0\leq\lambda_1\leq\lambda_2$ with:
\begin{equation} J_{ik}(t+\tau)=0\quad\mbox{if}\quad
  |s_i(t)-s_k(t)|=:|s_{ik}(t)|>\lambda_2\end{equation}
\begin{equation} J_{ik}(t+\tau)=\pm1\quad\mbox{if}\quad 0<\pm
  s_{ik}(t)<\lambda_1\end{equation}
with the special proviso that
\begin{equation} J_{ik}(t+\tau)=J_{ik}(t)\quad\mbox{if}\quad s_{ik}(t)=0
\end{equation}
On the other side
\begin{equation} J_{ik}(t+\tau)= \left\{\begin{array}{ll} 
\pm1 & \quad J_{ik}(t)\neq 0 \\
0    & \quad J_{ik}(t)=0
\end{array} \right. \quad\mbox{if}\quad
\lambda_1\leq\pm
  s_{ik}(t)\leq\lambda_2 
\end{equation}
In other words, bonds are switched off if local spatial charge
fluctuations are too large, switched on again if they are too
small, their orientation following the sign of local charge
differences, or remain inactive.
\end{defi}
Remarks:\hfill\\
\begin{enumerate}
\item It is important that, generically, such a law does not lead to a
reversible time evolution, i.e. there will typically exist attractors
in total phase space (the overall configuration space of
the node and bond states). On the other hand, there exist strategies
to develop reversible network laws (cf. e.g. \cite{12}).
\item In the above class of laws a direct bond-bond interaction is not
  yet implemented. We are prepared to incorporate such a (possibly
  important) contribution in a next step if it turns out to be
  necessary. In any case there are not so many ways to do this in a
  sensible way. Stated differently, the class of possible physically
  sensible interactions is perhaps not so numerous.
\item The above clock time should not be confused with so-called
  \tit{physical time}, which is, as is the case with many other
  physical continuum concepts, considered to be an emergent collective
  quantity, living on a higher level of the hierarchy. Note that its
  correct implementation is a big issue anyhow in quantum gravity and
  we refrain from overburding our paper with addressing this difficult
  topic (a recent review is e.g. \cite{Butter}). 
\item As in the definition of evolution laws of \tit{spin
  networks} by e.g. Markopoulou, Smolin and Borissov (see \cite{13} or
\cite{14}), there are in our case more or less two possibilities:
treating evolution laws within an integrated space-time formalism or
regard the network as representing space alone with the time evolution
being implanted via some extra principle ( which is the way we have
chosen above). The interrelation of these various approaches and
frameworks, while being very interesting, is however  far from obvious
at the moment and needs a separate detailed investigation.
\end{enumerate}
\begin{ob}[Gauge Invariance] The above dynamical law depends nowhere on the 
absolute values of the node charges but only on their relative
differences. By the same token, charge is nowhere created or destroyed. We have
\begin{equation}\sum_{QX}s(n)=\operatorname{const}\end{equation}
as far as the rhs is meaningful. One could e.g. choose as initial
condition respectively constraint 
\begin{equation}\sum_{QX}s(n)=0\end{equation} 
\end{ob}

There are many different aspects of our class of cellular network one
can study in this context. One can e.g. regard them as complex
dynamical systems, or one can decide to develop  a statistical or stochastic framework etc. In a purely
geometric sense, however, they are \tit{graphs}. On the other side, we are, at
least in this paper, primarily concerned with the analysis of the
\tit{microstructure} of (quantum) space-time. Therefore, in a first step, it
may be a sensible strategy to supress all the other features like e.g.
the details of the internal state spaces of nodes and bonds and
concentrate instead on their pure \tit{wiring diagram} and its reduced
(graph) dynamics. This is already an interesting characteristic of the
network (perhaps somewhat reminiscent of the \tit{Poincar\'e map} in
the theory of chaotic systems) as bonds can be switched on and off in
the course of clock time so that already the wiring diagram will
constantly change.  Furthermore, as we will see, it encodes the
complete \tit{near-} and \tit{far-order structure} of the network,
that is, it tells us which regions are experienced as near by or far
away (in a variety of possible physical ways such as strength of
correlations or with respect to some other physical metric like e.g.
\tit{statistical distance}) etc.  Evidently this is one of the crucial
features we expect from something like physical space-time. To give an
example. In the above simple scenario with $J_{ik}=\pm 1 $ or $0$ one
can e.g. draw a \tit{directed bond}, $d_{ik}$, if $J_{ik}=+1$, with
$J_{ik}=-J_{ki}$ implied, and delete the bond if $J_{ik}=0$. This leads
to a (clock) time dependent graph, $G(t)$, or \tit{wiring diagram}.

We close this section with a brief r\'esum\'{e} of the characteristics
an interesting network dynamics should encode (in our view).
\begin{res}
  Irrespectively of the technical details of the dynamical evolution
  law under discussion it should emulate the following, in our view
  crucial, principles, in order to match certain fundamental
  requirements concerning the capability of {\em emergent} and {\em
    complex} behavior.
\begin{enumerate}
\item As is the case with, say, gauge theory or general relativity,
  our evolution law on the surmised primordial level should implement
  the mutual interaction of two fundamental substructures, put
  sloppily: ``{\em geometry}'' acting on ``{\em matter}'' and vice
  versa, where in our context ``{\em geometry}'' is assumed to
  correspond in a loose sense with the local and/or global bond states
  and ``{\em matter}'' with the structure of the node states.
\item By the same token the alluded {\em selfreferential} dynamical
  circuitry of mutual interactions is expected to favor a kind of
  {\em undulating behavior} or {\em selfexcitation} above a return
  to some uninteresting `{\em equilibrium state}' as is frequently
  the case in systems consisting of a single component which directly
  acts back on itself. This propensity for the `{\em autonomous}'
  generation of undulation patterns is in our view an essential
  prerequisite for some form of ``{\em protoquantum behavior}'' we
  hope to recover on some coarse grained and less primordial level of
  the network dynamics.
\item In the same sense we expect the overall pattern of switched-on and
 -off bonds to generate a kind of ``{\em protogravity}''.
\end{enumerate}
\end{res}

\section{The Cellular Network as a (Random) Graph}

We start with the introduction of some graph theoretical definitions
(see e.g. \cite{11}).
\begin{defi}[Simple Locally Finite (Un)directed Graph)]\hfill
\begin{enumerate}
\item We write the {\em simple} {\em labeled} graph as $G:=(V,E)$ where $V$ is the
  countable set of {\em nodes} $\{n_i\}$ (or {\em vertices}) and $E$ the set of
  {\em bonds} ({\em edges}). The graph is called {\em simple} if there do not exist
  elementary {\em loops} and {\em multiple edges}, in other words: each
  existing bond connects two different nodes and there exists at most
  one bond between two nodes. (We could of course also discuss more
  general graphs). Furthermore, for simplicity, we assume the graph to
  be {\em connected}, i.e. two arbitrary nodes can be connected by a
  sequence of consecutive bonds called an {\em edge sequence} or {\em
    walk}. A
  minimal edge sequence, that is one with each intermediate node
  occurring only once, is called a {\em path} (note that these definitions
  may change from author to author).
\item We assume the graph to be {\em locally finite} (but this not
  always really necessary), that is, each node is incident with only a
  finite number of bonds. Sometimes it is useful to make the stronger
  assumption that this {\em vertex degree}, $v_i$, (number of bonds
  being incident with $n_i$), is globally bounded away from $\infty$.
\end{enumerate}
\end{defi}
\begin{obdef}
  Among the paths connecting two arbitrary nodes there exists at least
  one with minimal length. This length we denote by
  $d(n_i,n_k)$. This d has the properties of a metric, i.e:
\begin{eqnarray} d(n_i,n_i) & = & 0\\ d(n_i,n_k) & = &
d(n_k,n_i)  \\d(n_i,n_l) & \leq & d(n_i,n_k)+d(n_k,n_l) \end{eqnarray}
\end{obdef}
(The proof is more or less evident).
\begin{koro}
With the help of the metric one gets a natural
neighborhood structure around any given node, where ${\cal U}_m(n_i)$
comprises all the nodes, $n_k$, with $d(n_i,n_k)\leq m$, 
$\partial{\cal U}_m(n_i)$
the nodes with $d(n_i,n_k)=m$.
\end{koro}
Remark: The restriction to connected graphs is, for the time being,
only made for convenience. If one wants to study geometric phase
transitons of a more fragmented type, it is easy to include these more
general types of graphs. In the context of \tit{random graphs} ( which
we will introduce below) one can even derive probabilistic criteria
concerning geometric properties like connectedness etc.
\vspace{0.5cm}

We said above that, for the time being, we want to concentrate our
investigation on the geometric i.e. graph properties of our cellular
network which are themselves (clock) time dependent. On the other
hand, the graphs or networks we are actually interested in are
expected to be extremely large (number of nodes $\gtrapprox
10^{100}$). According to our philosophy they are to emulate the full
physical vaccum together with all its more or less macroscopic
excitations, or, in other words, the entire evolving universe.
Furthermore the assumed \tit{clocktime interval} $\tau$ is extremely
short (in fact of Plancktime order). On the other side, it is part of
our working philosophy that the phenomena we are observing in e.g.
present day high energy physics and, a fortiori, macroscopic physics
are of the nature of collective (frequently large scale) excitations
of this medium both with respect to space and time (in Planck units).
In other words, each of these patterns is expected to contain,
typically, a huge amount of nodes and bonds and to stretch over a
large number of clocktime intervals. This then suggests the following
approach which has been fruitful again and again in modern physics.
\begin{stat}
Following the above arguments, it makes sense to study so-called {\em
  graph properties} within a certain {\em statistical framework} to be
further explained below as long as one is interested in patterns which
are quasi macroscopic compared to the Planck scale.
\end{stat}
Remark: Similar strategies have been pursued in the investigation of
cellular automata (see e.g. \cite{Wolf} or \cite{Farm})
\\[0.5cm]
One strategy, i.e. errecting a \tit{probability space of graphs} over
the given set of nodes, will be introduced now.
\begin{random}Take all possible labeled graphs over $n$ nodes as
  probability space $\cal{G}$ (i.e. each graph represents an
  elementary event). The maximal possible number of bonds is
  $N:=\binom{n}{2}$, which corresponds to the unique {\em simplex
    graph} (denoted usually by $K_n$). Give each bond the {\em
    independent probability} $0\leq p\leq 1$, (more precisely, $p$ is
  the probability that there is a bond between the two nodes under
  discussion). Let $G_m$ be a graph over the above vertex set having
  $m$ bonds. Its probability is then
\begin{equation}pr(G_m)=p^m\cdot q^{N-m}\end{equation}
where $q:=1-p$. There exist $\binom{N}{m}$ different labeled
$m$-graphs $G_m$ and the above probability is correctly normalized,
i.e.
\begin{equation}pr({\cal G})=\sum_{m=0}^N {N\choose m}p^mq^{N-m}=(p+q)^N=1\end{equation}
This probability space is sometimes called the space of {\em
  binomially random graphs} and denoted by ${\cal G}(n,p)$. Note that
the number of edges is binomially distributed, i.e.
\begin{equation}pr(m)=\binom{N}{m}p^mq^{N-m}\end{equation}
and
\begin{equation}\langle m\rangle=\sum m\cdot pr(m)=N\cdot p\end{equation}
\end{random}
Proof of the latter statement:
\begin{equation}\langle
  m\rangle=d/d\lambda|_{\lambda=1}\left(\sum\binom{N}{m}(\lambda
    p)^mq^{N-m}\right)=d/d\lambda|_{\lambda=1}(\lambda
  p+q)^N=Np\end{equation}
or, with $e_i$ being the independent \tit{Bernoulli
  $(0,1)$-variables} (as to this notion cf. e.g. \cite{Feller}) belonging to the bonds:
\begin{equation}\langle e_i\rangle=p\quad\text{hence}\quad\langle
  m\rangle=\sum_1^N\langle e_i\rangle=Np \end{equation}
\hfill$\Box$\\
(The use of the above Bernoulli variables leads also to some
conceptual clarifications in other calculations). 
\begin{variant}A slightly different probability space can be
  constructed by considering only graphs with a fixed number, $m$, of
  bonds and give each the probability $\binom{N}{m}^{-1}$ as there are
  exactly $\binom{N}{m}$ of them. The corresponding probability space,
  $\mcal{G}(n,m)$, is called the space of {\em uniform random graphs}.
\end{variant}

The latter version is perhaps a little bit more common in pure mathematics
as this concept was introduced  mainly for purely combinatorial
reasons which have nothing to do with our own strand of ideas. The
whole theory was rather developed by Erd\"os and R\'enyi in the late
fifties and early sixties to cope with certain notorious (existence)
problems in graph theory (for more information see e.g. \cite{Random}
and \cite{Combi}, brief but concise accounts can also be found in
chapt.VII of \cite{11} or \cite{Kar}).
\begin{ob}The two random graph models behave similarly if $m\approx
  p\cdot N$. Note however, that there exists a subtle difference
  between the two models anyway. In the former model all elementary
  bond random variables are {\em independent} while in the latter case
  they are (typically weakly) dependent.
\end{ob}
(While being plausible this statement needs a proof which can be found
in e.g. \cite{Random}).\vspace{0.5cm}

The really fundamental observation made already by Erd\"os and R\'enyi (a
rigorous proof of this deep result can e.g. be found in \cite{Combi})
is that there are what physicists would call \tit{phase transitions}
in these \tit{random graphs}. To go a little bit more into the details
we have to introduce some more graph concepts.
\begin{defi}[Graph Properties]{\em Graph properties} are certain
  particular {\em random
    variables} (indicator functions of so-called events) on the above
  probability space ${\cal G}$. I.e., a graph property, $Q$, is
  represented by the subset of graphs of the sample space having the
  property under discussion. To give some
  examples: i) connectedness of the graph, ii) existence and number of certain particular subgraphs (such as
  subsimplices etc.), iii) other geometric or topological graph
  properties etc.
\end{defi}
Remark: In addition to that there are other more general random
variables (\tit{`graph characteristics'}) describing the fine
structure of graphs, some of which we will introduce
below.\vspace{0.5cm}

In this context Erd\"os and R\'enyi made the following important
observation.
\begin{ob}[Threshold Function]A large class of {\em graph properties}
  (e.g. the {\em monotone increasing ones}, cf. the above cited
  literature) have a so-called {\em threshold function}, $m^*(n)$,
  so that for $n\to\infty$ the graphs under discussion have {\em
    property} $Q$ {\em almost shurely} for $m(n)>m^*(n)$ and {\em
    almost shurely not} for $m(n)<m^*(n)$ or vice versa (more
  precisely: for $m(n)/m^*(n)\to \infty\;\text{or}\;0$; for the
  details see the above literature). The above version applies to the
  second kind of graph probability space, ${\cal G}(n,m)$. A
  corresponding result holds for ${\cal G}(n,p)$ with $p(n)$ replacing
  $m(n)$. That is, by turning on the probability $p$, one can drive
  the graph one is interested in beyond the phase transition threshold
  belonging to the graph property under study. Note that, by
  definition, threshold functions are only unique up to
  ``factorization'', i.e. $m^*_2(n)= O(m^*_1(n))$ is also a
  threshold function.
\end{ob}
\begin{bsp}[Connectedness] \label{connectedness}The threshold function for the graph
  property {\em connectedness} is 
\begin{equation}m^*(n)=n/2\cdot\log(n)\;\text{respectively}\;
  p^*(n)=\log(n)/n\end{equation}
Note that with the help of the above observation, i.e. for $m\approx
p\cdot\binom{n}{2}$, we have for $n$ large: $\binom{n}{2}\approx
n^2/2$ and hence $p\cdot n^2/2\approx n\log(n)$, i.e. $p\approx \log(n)/n$
\end{bsp}

In the following our main thrust will go towards the developement of
the concept of \tit{proto space-time} as an \tit{order parameter
  manifold} or \tit{superstructure} floating in our network $QX$
and the concept of \tit{physical points}. We therefore illustrate the method of random graphs, graph
properties and graph characteristics by applying it to a particular feature being of importance
in the sequel.
\begin{defi}[Subsimplices and Cliques]With $G$ a given fixed graph and
  $V_i$ a subset of its vertex set $V$, the corresponding {\em induced
    subgraph} over $V_i$ is called a subsimplex (ss),
  $S_i$, or {\em complete subgraph}, if all its nodes are connected by
  a bond. In this class there exist certain {\em maximal subsimplices}
   $(mss)$, that is, every addition of another node destroys this
  property. These $mss$ are called {\em cliques} in combinatorics and
  are the candidates for our {\em physicaal points} carrying a
  presumably rich internal (dynamical) structure.
\end{defi}
We consider all possible graphs, $G$, over the fixed vertex set $V$ of
$n$ nodes. For each subset $V_i\subset V$ of order $r$ (i.e. number of
elements) we define the following random variable:
\begin{equation}X_i(G):=
\begin{cases}1 & \text{if $G_i$ is an $r$-simplex},\\  
 0 & \text{else}
\end{cases}
\end{equation}
where $G_i$ is the corresponding induced subgraph over $V_i$ with
respect to $G\in {\cal G}$ (the probability space). Another random
variable is then the \tit{number of $r$-simplices in $G$}, denoted by
$Y_r(G)$ and we have:
\begin{equation}Y_r=\sum_{i=1}^{\binom{n}{r}}X_i\end{equation}
with $\binom{n}{r}$ the number of $r$-subsets $V_i\subset V$. With respect
to the probability measure introduced above we then have for the
\tit{expectation values}:
\begin{equation}\langle Y_r \rangle = \sum_i \langle X_i \rangle\end{equation}
and
\begin{equation}\langle X_i \rangle = \sum_{G\in{\cal G}} X_i(G)\cdot
  pr(G_i=\text{$r$-simplex in}\;G).\end{equation} 
With the sum running
over all $G\in {\cal G}$ and $X_i$ being one or zero we get
\begin{equation}\langle X_i \rangle = pr(G_i\;\text{an $r$-simplex,
    $G-E_i$ an arbitrary graph})\end{equation}
where $G-E_i$ is the remaining graph after all the edges belonging to
$G_i$ have been eliminated. This yields 
\begin{equation}\langle X_i \rangle =
  p^{\binom{r}{2}}\cdot\sum_{G'\in{\cal G}'} pr(G')\end{equation}
where ${\cal G}'$ is the probability space of graphs over $V$ with all
the bonds $E_i$ being omitted. The maximal possible number of bonds
belonging to ${\cal G}'$ is
\begin{equation}|E'|=\binom{n}{2}-\binom{r}{2}.\end{equation}
Each of these bonds can be on or off with probability $p$ or
$(1-p)$. To each graph of ${\cal G}'$ belongs a unique labeled
sequence of $p$'s and $q$'s and every such sequence does occur
(i.e. with either $p$ or $q$ at label $i$). We hence have
\begin{equation}\sum_{G'}pr(G')=(p+q)^{|E'|}=1^{|E'|}=1\end{equation}
and we get
\begin{equation}\langle X_i \rangle = p^{\binom{r}{2}}\end{equation}
The case that such a $r$-simplex is already maximal, i.e. is actually
a clique, can be calculated as follows. The addition of a single
further vertex would destroy the property of being a $ss$.  In other
words, each of the vertices in the graph $G$, not lying in the
node-set $V_i$, is connected with the above vertex set, $V_r$, by
fewer than $r$ bonds. This can now be quantitatively encoded as
follows: The probability that the induced subgraph $G_r$ over $r$
arbitrarily chosen vertices is already a $mss$ is the product of the
{\em independent} probabilities that it is a $ss$ and that {\em each}
of the remaining $(n-r)$ vertices is connected with $V_r$ by fewer
than $r$ bonds. The latter probability is $(1-p^r)^{n-r}$, hence
\begin{equation}pr(G_r\;\text{is a clique})=(1-p^r)^{n-r}\cdot
  p^{\binom{r}{2}}\end{equation} 
There are now exactly $\binom{n}{r}$ possible $r$-subsimplices over
the node set $V$. We arrive therefore at the following important conclusion: 

\begin{conclusion}[Distribution of Subsimplices and Cliques]The
  expectation value of the random variable {\em number of $r$-subsimplices} is
\begin{equation}\langle Y_r \rangle = \binom{n}{r}\cdot
  p^{\binom{r}{2}}\end{equation}
 For $Z_r$, the number of $r$-cliques in the random graph, we have
 then the following relation
\begin{equation}\langle Z_r
  \rangle=\binom{n}{r}\cdot(1-p^r)^{n-r}\cdot
  p^{\binom{r}{2}}\end{equation}
\end{conclusion}
It is remarkable, both physically and combinatorially, that these
quantities, as  functions of $r$ (the \tit{order} of subsimplices) have
quite a peculiar numerical behavior. We are, among other things,
interested in the typical order of cliques (where typical is
understood in a probabilistic sense).
\begin{obdef}[Clique Number]The maximal order of  cliques contained in
  $G$ is called its {\em clique number}, $cl(G)$. It is another random
  variable on the probability space ${\cal G}(n,p)$.
\end{obdef}
An analysis of the above expression $\langle
Y_r\rangle=\binom{n}{r}\cdot p^{\binom{r}{2}}$ as a function of $r$
shows that it is typically very large (for $n$ sufficiently large) for
all $r$ lying in a certain interval below some \tit{critical value},
$r_0$, and drops rapidly to zero for $r> r_0$.
\begin{conclusion}From the above one can infer that this value $r_0$
  is a good approximation for the above defined {\em clique number} of a
  typical random graph, depending only on $p$ and $n$. In other words,
  it approximates the order of the largest occurring cliques in a
  typical random graph. $r_0$ can be approximated as
\begin{equation}r_0\approx 2\log(n)/\log(p^{-1})+ O(\log\log(n))\end{equation} 
{\em(cf. chapt. XI.1 of \cite{Random}).}
\end{conclusion}
Remark: A numerical analysis of the geometric web of cliques and
various of its characteristics is given in sect.4.\\[0.5cm]
\begin{conclusion}By an estimation of the {\em variance} of $Z_r$ we
  can conclude that the typical orders of occurring cliques lie
  between $r_0(n)/2$ and $r_0(n)$.
\end{conclusion}
\begin{bem}To make the above reasoning perhaps more transparent,
  it is again helpful to exploit the properties of the elementary
  $(0,1)$-edge-variables $e_i$. The probability that $r$ arbitrarily
  selected bonds exist in the random graph is
  $pr(e_1=\ldots=e_r=1)=p^r$, the complimentary possibility (i.e.,
  that some of these bonds are missing), has hence the probability
  $(1-p^r)$.
\end{bem}
\begin{res}[Random Graph Approach]There is of course no absolute
  guarantee that our network, following a {\em deterministic}
  evolution law and typically reaching after a certain {\em transient}
  one of possibly several attractors, can in every respect be regarded
  as the evolution of true random graphs. In other words, its behavior
  cannot be entirely random already by definition. Quite to the
  contrary, we expect a shrinking of the huge accessible {\em phase
    space} during its evolution which manifests itself on a more
  macroscopic level as {\em pattern creation}.
  
  The underlying strategy is rather the following. At each clock time
  step, $G(t)$ is a graph having a definite number of active bonds,
  say $m$.  Surmising that $G(t)$ is sufficiently generic or typical
  we may be allowed to regard it as a typical member of the
  corresponding family ${\cal G}(n,m)$ or ${\cal G}(n,p)$ (at least as
  far as certain gross features are concerned). Via this line of
  inference we are quite confident of being able to get some clues
  concerning the qualitative behavior of our network, the microscopic
  time evolution of which is too erratic to follow in detail. As to
  this working philosophy the situation is not so different from the
  state of affairs in many areas of, say, ordinary statistical physics
  or complex dynamical systems . But nevertheless, a more detailed
  analysis (underpinned by concrete numerical simulations) of the
  validity or limits of such an ansatz would be desirable but has to
  be postponed in order not to blow up the paper to much (cf. e.g. the
  completely similar situation in the context of \tit{cellular
    automata}, being expounded to some extent in the above mentioned
  literature).
  \end{res}
  Some steps in this direction were taken by our former student Th.
  Nowotny as part of his diploma thesis. Other perhaps characteristic numerical deviations between theoretical
  results based on certain apriori statistical assumptions and
  computer simulations of concrete models were also found in
  \cite{Antonsen} by Antonsen, who presented an approach which is
  different from ours in several respects but belongs to the same
  general context.
\section{The Emergence of (Proto) Space-Time}
In this core section of our investigation we are going to describe how
the presumed underlying (discrete) fine structure of our continuous
space-time may look like. We want however to emphasize at this place
that, while most of the details and observations we will present are
rigorously proved, the overall picture is still only
hypothetical. That is, we are able to describe a, as we think, fairly
interesting scenario in quite some detail but are not yet able to show
convincingly that our network actually evolves into exactly such a
phase we are going to expound in the following under one of the
dynamical laws we described above. We will in fact show, that the
unfolding of something like an extended, macroscopically stable
space-time structure having, afortiori, an integer macroscopic
dimension, is quite a subtle phenomenon from a purely probabilistic
point of view and that only the understanding of the nature of these
subtleties will lead us to the correct class of dynamical laws. 

The central theme of this paper is the description and analysis of a
certain superstructure, $ST$, emerging within our network $QX$ as a
consequence of a process which can be interpreted as a geometrical
phase transition. In this picture $ST$, which we experience as
space-time, plays the role of an order parameter manifold. Its
emergence signals the transition from a \tit{disordered} and
\tit{chaotic} initial phase to a phase developing a
\tit{near-/far-order}, i.e. a \tit{causal structure}, and stable
\tit{physical points} or \tit{lumps} (Menger).

Our qualitative picture concerning the initial scenario is the
following (more details can be found in e.g. section 4 of
\cite{Planck}). The network, $QX$, started from a presumed densely
entangled \tit{initial phase}, $QX_0$, in which on average every pair
of nodes was connected by an active bond with high probability
(i.e. it was almost a simplex). This chaotic initial phase was
characterized by extremely large fluctuations of local node/bond
charges and very short-lived/short-ranged correlations. This is a
result of the dense entanglement and the kind of network laws we
described in sect.2 (they typically favor to some extent an
\tit{overshooting}). There are, in particular, no stable cliques or
lumps and no traces of a kind of proto space-time.

We conjecture now that, under certain favorable conditions, that is,
an appropriately chosen dynamical law together with the occurrence of
a suitably large collective spontaneous deviation of the network state
in some region, the network is capable of leaving this chaotic initial
phase, starting from such a \tit{nucleation center}, and unfolds
towards a new more ordered attractor, the above mentioned phase
$QX/ST$. This geometric phase transition is triggered respectively
accompanied by an annihilation of a certain fraction of active bonds
(in the language of our network laws: $J_{ik}=0$) due to too large
differences in neighboring node charges. In this process the
individual and incoherent elementary fluctuations are expected to be
reorganized in a more macroscopic pattern (in the language of
\tit{synergetics} they become \tit{slaved}; see e.g. \cite{Haken}). We
note in passing that what we indicated above may be described in a
more conventional context as the \tit{big bang scenario} and it
remains  a big task to fill out all the intermediate steps.

We now describe in broad outline our idea of the underlying discrete
substratum of space-time and relate it, in a next step, to the more
rigorously defined mathematical correlatives within the random graph
framework. We want to emphasize that, for the time being, the only
structural or quantitative input we are employing to encode the
effects of the presumed geometric phase transition is the decrease in
the number of occupied bonds in the wiring diagram of the
network. That is, we will mainly be occupied with the description and
analysis of the structural and geometric changes, occurring in the
underlying graph, when the \tit{bond probability}, $p(t)$, decreases
with $t$ the \tit{clock time}. 
\begin{points}[Physical Points]\hfill
\begin{enumerate}
\item Physical points have a (presumably rich) internal structure,
  i.e. they consist of a (presumably) large number of nodes and
  bonds. In the words of Menger they are lumps.
\item We suppose that, what we are used to decribe as fields at a
  space-time point (in fact, rather {\em distributions} in
  e.g. quantum field theory), are really internal excitations of these
  lumps.
\item In order to have a qualitative measure to tell the physical points apart, that is, to discern what happens within a certain point or
  between different points, we identify the physical points with  particularly densely connected subgraphs of our network
  or graph. This then motivates our interest in maximal subsimplices
  or cliques as natural candidates.
\item Typically (i.e. if a certain fraction of bonds has been
  eliminated), some of these lumps overlap with each other in a
  stronger or weaker sense, forming so to say {\em local groups}
  while other will cease to overlap. This will then establish a kind
  of {\em proto-causality} or near/far-order in our {\em proto-space-time} and will be one of the central topics in our analysis.
\end{enumerate}
\end{points}
To get an idea how the \tit{order} of the typical cliques depends on
the bond probability, $p$, we apply the formula for the \tit{clique
  number}, provided in sect.3 to a graph having, say, $10^{100}$ nodes
with $p$ varying (we omitted the $log(log(n))$-term).
\begin{equation}\label{tabelle1}
\begin{array}{|l|c|c|c|c|c|c|c|c|c|}
p & 0.9 & 0.8 & 0.7 & 0.6 & 0.5 & 0.4 & 0.3 & 0.2 & 0.1\\ \hline
r_0 & 4444 & 2083 & 1333 & 909 & 666 & 500 & 400 & 285 & 200
\end{array}
\end{equation} 
We will proceed by compiling a couple of simple observations
concerning $(m)ss$, proofs being partly omitted (if they are obvious).
\begin{ob}\hfill
\begin{enumerate}
\item If the node degree, $v_i$, of $n_i$ is
  smaller than $\infty$ then $n_i$ can lie in at most a finite set of
  different simplices, an upper bound being provided by the number of
  different subsets of bonds emerging from $n_i$, that is $2^{v_i}$.
\item The set of subsimplices is evidently {\em partially ordered} by
  inclusion
\item Furthermore, if S is a simplex, each of its subsets is again a
  simplex (called a {\em face})
\item It follows that each of the {\em chains} of linearly ordered
  simplices (containing a certain fixed node) is finite. The corresponding length can be calculated in a
  similar way as in item 1 by selecting chains of sets of bonds,
  ordered by inclusion. In
  other words each chain has a maximal element. By the same token each
  node lies in at least one (but generically several) mss
\item A $mss$ with $n_i$ being a member can comprise at most $(v_i+1)$
  nodes, in other words, its order is bounded by the minimum of these
  numbers when $n_i$ varies over the mss.
\end{enumerate}
\end{ob}
Proof of item 1: Assume that $S_k,S_l$ are two different simplices containing
$n_i$. By definition $n_i$ is linked with all the other nodes in $S_k$
or $S_l$. As these sets are different by assumption, the corresponding
subsets of bonds emerging from $n_i$ are different. On the other side,
not every subset of such bonds corresponds to a simplex (there
respective endpoints need not form a simplex), which proves the upper
bound stated above.\hfill$\Box$
\begin{ob}The class of simplices, in particular
  the mss, containing a certain fixed node, $n_i$, can be generated in
  a completely algorithmic way, starting from $n_i$. The first level
  consists of the bonds with $n_i$ an end node, the second level
  comprises the triples of nodes ({\em triangles}), $(n_in_kn_l)$, with
  the nodes linked with each other and so forth. Each {\em level set} can be
  constructed from the preceding one and the process stops when a mss
  is reached.
\end{ob}
Remark: Note that at each intermediate step, i.e. having
already constructed a certain particular subsimplex, one has in general
several possibilities to proceed. On the other hand, a chain of such
choices may differ at certain places from another one but may lead in
the end to the same final simplex (in other words, being simply a permutation of the nodes of the
former simplex).\vspace{0.5cm}

Denoting the $(m)ss$ under discussion by capital $S$ with
certain indices or labels attached to it, this process can be
pictorially abreviated as follows:
\begin{equation}S(n_0\to\cdots\to n_k)\end{equation}
With $S(n_0\to\cdots\to n_k)$ given, each permutation will yield the
same $mss$, i.e:
\begin{equation}S(n_0\to\cdots\to n_k)=S(n_{\pi(0)}\to\cdots\to n_{\pi
    (k)})\end{equation}
Furthermore each $mss$ can be constructed in this way, starting from one of its
nodes. Evidently this could be done for each node and for all possible
alternatives as to the choice of the next node in the above sequence.
\begin{defi} Let $G_{\nu}$ be a class of subgraphs of $G$.\\
\begin{enumerate}
\item $\cap G_{\nu}$ is the graph with $n\in V_{\cap G_{\nu}}$ if $n\in$
every $V_{G_{\nu}}$,\\
$b_{ik}\in E_{\cap G_{\nu}}$ if $b_{ik}\in$ every $E_{G_{\nu}}$\\
\item $\cup G_{\nu}$ is the graph with $n\in V_{\cup G_{\nu}}$ if $n\in
V_{G_{\nu}}$ for at least one $\nu$\\
$b_{ik}\in E_{\cup G_{\nu}}$ if $b_{ik}\in E_{G_{\nu}}$ for at least
one $\nu$.
\end{enumerate}
\end{defi}
As every node or bond belongs to at least one $mss$ (as can be easily
inferred from the above algorithmic construction), we have
\begin{koro}\begin{equation}\cup S_{\nu}=G\end{equation}
\end{koro}

After these preliminary remarks we now turn to our main task, that is,
the analysis of the web of these $mss$ as the elementary building
blocks of the next higher level of organisation. 
\subsection{The Embryonic Epoch}
In its surmised transition from the almost maximally connected and
chaotic initial phase to the fully developed phase, $QX/ST$ (i.e. $QX$
plus superstructure $ST$), the
underlying graph passes through several clearly distinguishable
epochs. In the following we study two main phases of the network. The
first one, which we call embryonic, is characterized by a still
pronounced common overlap of the emerging lumps or physical
points. That is, they can still interact directly with each other
which implies that there exist no true far-order or larger distances
on the network of lumps. This epoch should be typical (in our picture)
for the infinitesimal time interval just after the \tit{big bang}.

 We begin with the epoch where only a small fraction of bonds
is shut off. Let us e.g. assume that $\alpha$ bonds with  
\begin{equation}1\ll\alpha<n/2 \ll n(n-1)/2=N\quad\text{for $n$ large}\end{equation}
are temporarily dead with $n$ the order of the graph or network
(number of nodes) and $N$ the maximal possible number of bonds. In
other words, the network is supposed to be still near the initial
phase. We observe that $\alpha$ arbitrarily selected bonds can at most
connect $k\le 2\alpha$ different nodes, hence there still exist at
least $(n-k)$ nodes which are maximally connected, viz. they are
spanning a still huge subsimplex $S'\subset G$. On the other hand
there are at most $k\le 2\alpha$ nodes with one or more incident bonds
missing in the corresponding induced subgraph.

$V_G$ can hence be split in the following way:
\begin{equation} V_G=V_{S'}\cup V_N\end{equation}
with $V_N$ the unique set of those nodes so that to each node $n_i$ in $V_N$
there exist other nodes (also lying in $V_N$) with the respective
bonds to $n_i$ missing. $V_{S'}$, on the other side, is the set of
remaining nodes which are maximally connected (by construction); i.e. they form a $ss$.
\begin{equation} |V_N|=k\leq 2\alpha\end{equation}
\begin{defi}$[\cup G_i]$ is the induced subgraph spanned by the nodes
  occurring   in $\cup V_i$. Note that in general $[\cup G_i]\supset
  \cup G_i$, that is, it may rather be called its {\em `closure'}.
\end{defi}
\begin{ob} \label{Durchschnitt}       \hfill
\begin{enumerate}
\item The simplex $S'$ is contained in each of
the occurring $mss$, $S_{\nu}$, i.e:
\begin{equation} S'\subset\cap S_{\nu}\;\text{and it holds a fortiori}\;S'=\cap
S_{\nu}\end{equation}
We hence have 
\begin{equation}|V_{S'}|=|V_{\cap S_{\nu}}\ge n-2\alpha\end{equation}
\item Note that $S'$ itself is never maximal since $[S'\cup n_i]$ is always a larger
simplex with $n_i\in N$ and $[S'\cup n_i]$ being the induced subgraph
spanned by $V_{S'}$ and $n_i$.
\item To each maximal simplex $S_{\nu}\subset G$ belongs a unique
maximal subsimplex $N_{\nu}\subset N$ with
\begin{equation} S_{\nu}=[S'\cup N_{\nu}]\end{equation} 
\item It is important for what follows that $S'$ can be uniquely
  characterized, without actually knowing the $S_{\nu}$, by the
  following two properties: i)$S'$ is a $ss$ so that all bonds
  connecting nodes from $V_{S'}$ with $V-V_{S'}$ are ``on''. ii) $S'$ is
  maximal in this class of $ss$, that is, each node in $V-V_{S'}$ has
  at least one bond missing with respect to the other nodes in
  $V-V_{S'}$. An induced subgraph in $G$, having these properties is
  automatically the uniquely given $S'$!
\end{enumerate}
\end{ob}
\begin{koro} From the maximality of the $N_{\nu}$ follows a
general structure relation for the $\{S_{\nu}\}$ and $\{N_{\nu}\}$:
\begin{equation} \nu\neq\mu\,\to\,S_{\nu}\neq S_{\mu}\,\to\,N_{\nu}\neq
N_{\mu}\end{equation}
and neither
\begin{equation} N_{\nu}\subset N_{\mu}\;\mbox{nor}\;N_{\mu}\subset N_{\nu}\end{equation}
viz. there always exists at least one $n_{\nu}\in V_{N_{\nu}}$
s.t. $n_{\nu}\notin V_{N_{\mu}}$ and vice versa.
\end{koro}
Proof of the above observation:
\begin{enumerate}
\item Starting from an arbitrary node $n\in
G$, it is by definition connected with all the nodes in $S'$,
since if say $n,\,n'$ are not connected they both belong to $N$ (by
definition). I.e., irrespectively how we will proceed in the
construction of some $S_{\nu}$, $S'$ can always be added at any
intermediate step, hence $S'\subset\cap S_{\nu}$. On the other side
assume that $n\in \cap S_{\nu}$. This implies that $n$ is connected
with each node in $\cup S_{\nu}$. We showed above that $\cup
S_{\nu}=G$, hence $n$ is connected with all the other nodes, i.e. it
is not in $N$, that is, $n\in S'$, which proves the statement.
\item  As $n\in N$ is connected with each $n'\in S'$ (by definition of
$N$ and $S'$), the subgraph $[S'\cup n]$ is again a (larger)
simplex.
\item We have $S'\subset S_{\nu}$ for all $\nu$, hence 
\begin{equation} S_{\nu}\neq S_{\mu}\;\mbox{implies}\;N_{\nu}\neq N_{\mu}\end{equation}
with $N_{\nu,\mu}$ the corresponding subgraphs in $N$.\\
With $S_{\nu}$ being a simplex, $N_{\nu}$ is again a subsimplex which
is maximal in $N$. Otherwise $S_{\nu}$ would not be maximal in $G$.\\
On the other side each $S_{\nu}=[S'\cup N_{\nu}]$ is uniquely given by
a maximal $N_{\nu}$ in $N$ as each node in $N$ is connected with all
the nodes in $S'$.\end{enumerate}\hfill$\Box$
\vspace{0.5cm}

We see from the above that as long as $\alpha$, the number of dead
(missing) bonds, is sufficiently small, i.e. $2\alpha<n$, there does
necessarily exist a non-void overlap $S'$, among the class of $mss$,
$S_{\nu}$. This overlap will become smaller as $\alpha$ increases. By
the same token the number of $mss$ will increase for a certain range
of the parameter $\alpha$ while the respective size of the $mss$ will
shrink. The above results hold for each given graph. On the other
side, the above unique characterization of $S'$ in item 4 makes it
possible to attack the problem of the order of $S'$ within the
framework of \tit{random graphs} in a more quantitative manner. Given a
member $G$ of ${\cal G}(n,p)$, $S'$ is fixed by item 4 of the above
observation. We are interested in the probability of $S'$ having, say,
$r$ nodes.

The strategy is, as usual, to try to express the probability of such a
configuration within $G$ as the product of certain more elementary and
(if possible) independent probabilities. Unfortunately this turns out
to be relatively intricate in the above case and we are, at the
moment, only able to provide certain upper and lower bounds for the
probability under discussion. As this example shows that such
questions may not always have simple and straightforward answers, it
is perhaps worthwhile to dwell a little bit on this point.

The typical difficulties one usually encounters in this context are
the following: The structure of the set of graphs in ${\cal G}(n,p)$
having a prescribed property may be rather complicated, so that it is
difficult to avoid multiple counting of members when trying to
calculate the order of such a set. A frequent reason for this is the
intricate entanglement of the various pieces of a complicated graph, a
case in point being the above description of $S'$ in $G$. In our case
the peculiar entanglement can be seen as follows.

Selecting $r$ arbitrary vertices, the probability that the
corresponding induced subgraph forms a $ss$ is $p^{\binom{r}{2}}$ (see
section 3). If this subgraph is to qualify as $S'$, i.e. $S'=\cap
S_{\nu}$, $\cup S_{\nu}=G$, each of the nodes in $N$ is connected with
every node in $S'$. The probability for this property is
$p^{r\cdot(n-r)}$. The difficult part of the reasoning concerns the
subgraph $N$. We call the probability that $N$ has just the structure
being described above, $pr(N)$.

The following observation is helpful. As $S'$ is unique in $G$, i.e.
occurs only once, the corresponding random variable, $X_{S'_r}$, that we
hit at such an $S'$ of order $r$, when browsing through the set of
induced $r$-subgraphs, is zero with at most one possible
exception, that is, if $S'$ has just the order $r$. Therefore the
corresponding expectation value of $X_{S'_r}$ is, by the same token,
also the probability of the property ($S'_r$).
\begin{conclusion} The probability that a random graph contains such a
 $S'$ of order r is
\begin{equation}pr(S'_r)=\binom{n}{r}\cdot p^{\binom{r}{2}}\cdot
  p^{r(n-r)}\cdot pr(N)\end{equation}
\end{conclusion}
As far as we can see, it is not easy to disentangle $pr(N)$ into more
elementary \tit{independent} probabilities and master the complex
combinatorics. Therefore we will, at the moment, only give (possibly
crude) upper and lower bounds. 

$N$, having $(n-r)$ nodes, is characterized as follows. Labeling the
nodes from $(1)$ to $(n-r)$, none of them is allowed to have the
maximal possible degree (with respect to $N$), i.e. $(n-r-1)$. The
first step is simple. Starting with, say, node $(1)$, the probability
that at least one bond is missing is the complement of the probability
that all possible bonds are present, i.e. $(1-p^{(n-r-1)})$. The
following steps will however become more and more cumbersome. Take
e.g. node $(2)$. In the above probability is already contained both
the probability that either the bond $b_{12}$ is missing or not. If
$b_{12}$ is not missing then some other bond $b_{1i}$ has to be absent
(by the definition of $N$). These two alternatives influence the
possible choices being made at step two. In the former case the
configuration where all bonds $b_{2j},\;j>2$ are present is
admissible, in the latter case this possibility is forbidden.
Depending of which choice we make at each step the algorithmic
construction bifurcates in a somewhat involved manner. Evidently this
first step yields a crude upper bound on $pr(N)$. Making at each step
$(i)$ the particular choice that there are always missing bonds among
the bonds pointing to nodes $(j)$ with $j>i$ provides a lower bound.
We hence have.
\begin{conclusion}
\begin{equation}(1-p^{n-r-1})\geq
  pr(N)\geq\prod_{j=1}^{n-r-1}(1-p^{n-r-j})=\prod_{j=1}^{n-r-1}(1-p^j)\end{equation}
and for $pr(S'=\emptyset)$:
\begin{equation}(1-p^{n-1})\geq
  pr(S'=\emptyset)\geq\prod_{j=1}^{n-1}(1-p^j)\end{equation}
\end{conclusion}
\begin{ob}The lower bound is interesting! Perhaps surprisingly, the
  occurring product is an important {\em number theoretic function}
  belonging to the field of {\em partitions of natural numbers} (see
  any good textbook about combinatorics or the
  famous book of Hardy and Wright, \cite{Hardy}, the standard source
  being \cite{Andrews}). Our above random graph approach offers the
  opportunity to (re)derive and prove this number theoretic formula by
  purely probabilistic means, i.e. give it an underpinning which seems
  to be, at first glance, quite foreign. We will come back to this
  interesting point elsewhere.
\end{ob}
It is important to have effective estimates for the regime of
probabilities, $p(n)$, so that $S'$ is empty with a high probability.
According to our philosophy this signals the end of the \tit{embryonic
  epoch}, where all the supposed \tit{protopoints} still overlap (and
hence are capable of direct interaction) and the beginning of the
unfolding of a new phase with, as we hope, a more pronounced
\tit{near- and far-order} among the physical points.

Such an estimate can in fact be provided with the help of the above
inequality. We have
\begin{equation}pr(S'=\emptyset)>\prod_1^{n-1}(1-p^j)>\prod_1^{\infty}(1-p^j)=\sum_{k=0}^{\infty}a_kp^k\end{equation}
for $0<p<1$. The following (highly nontrivial) observation is due to
Euler (cf. the above mentioned literature for more recent proofs):
\begin{satz}
\begin{multline}\prod_1^{\infty}(1-p^j)=\sum_{k=0}^{\infty}(-1)^k\cdot\left(p^{\frac{1}{2}(3k^2+k)}+p^{\frac{1}{2}(3k^2-k)}\right)=1-p-p^2+p^5+p^7\ldots\\
\approx
  1-p-p^2\end{multline}
for $p$ small.
\end{satz}
\begin{conclusion}For p near zero, $S'$ is empty with arbitrarily large
  probability $\lesssim 1$.
\end{conclusion}
This shows that there exists in fact a regime of small $p$-values
where the embryonic epoch no longer prevails. This holds the more so
for an $n$-dependent $p$ (which is very natural) and $p(n)\searrow 0$.
On the other side, note that there exists a possibly substantial class
of bond configurations which have, up to now, been excluded in the
above estimate, the inclusion of which would increase the relevant
probability further.

To get a feeling how good the above exact estimate may be, we
construct a (as we think, not untypical) example. The construction
goes as follows. Take $2k$ nodes, choose a subset $G_1$ consisting of
exactly $k$ nodes $(n_1.\ldots,n_k)$, make $G_1$ a simplex. With the
remaining $k$ nodes $(n_1',\ldots,n_k')$ we proceed in the same way,
i.e. we now have two subsimplices $G_1,G_1'$. We now choose a
one-one-map from $(n_1,\ldots,n_k)$ to $(n_1',\ldots,n_k')$, say:
\begin{equation}n_i\;\to\;n_i'\end{equation}
We now connect all the $n_i$ with the $n_j'$ except for the $k$ pairs
$(n_i,n_i')$. The graph $G$ so constructed has 
\begin{equation}|E_G|=2k(2k-1)/2-k=2k(2k-2)/2\end{equation}
We see from this that, as in our above network scenario, the number of
missing bonds is a relatively small fraction. We can now make the
following sequence of (easy to prove) observations:
\begin{ob} $G_1=S_1$ is already a $mss$ as each
  $n_i'\in G_1'$ has one bond missing with respect to $G_1$. One gets
  new $mss$ by exchanging exactly one $n_i$ with its partner $n_i'$,
  pictorially:
\begin{equation}[S_1-n_i+n_i']\end{equation}
yielding $k$ further $mss$. One can proceed by constructing another
class of $mss$, now deleting $(n_i,n_j)$ and adding their respective
partners, i.e:
\begin{equation}[S_1-n_i-n_j+n_i'+n_j']\end
{equation}
This can be done until we end up with the $mss$
\begin{equation}[S_1-n_1-\cdots-n_k+n_1'+\cdots+n_k']=S_1'\end{equation}
The combinatorics goes as follows:
\begin{equation}|\{mss\}|=\sum_{\nu=0}^k {k \choose
    \nu}=(1+1)^k=2^k\end{equation}
i.e., our $2k-$node-graph (with $k$ bonds missing) contains exactly
$2^k$ $mss$ of order $k$. Evidently $S'=\emptyset$
\end{ob}

We showed above that $S'$ is non-empty as long as $\alpha$, the number
of missing bonds, is smaller than $n/2$, $n$ the order of the graph
$G$, since $\alpha$ bonds can at most connect $2\alpha$ different
nodes. On the other side, $\alpha=n/2$ implies
\begin{equation}|E_G|=\binom{n}{2}-n/2=n(n-2)/2\end{equation}
or an \tit{average vertex degree}
\begin{equation}\langle v(n)\rangle_s=n-2\end{equation}
which is still very large. The example constructed in the preceeding observation has
\begin{equation}n=2k\;,\;\alpha=k=n/2 \;,\;S'=\emptyset\end{equation}
In other words, the parameter $\alpha$ is just the \tit{critical} one
given in \ref{Durchschnitt}. One may hence surmise that
  $\alpha=n/2$ is perhaps the threshold for $S'=\emptyset$ in the
  sense that, say, 
\begin{equation}pr(S'=\emptyset)=O(1)\end{equation}
for $\alpha\gtrsim n/2$. 

On the other side, within the framework of random graphs, we have
obtained the rigorous but presumably not optimal estimate
\begin{equation}pr(S'=\emptyset)>\prod_1^{\infty}(1-p^j)\approx
  1-p-p^2\end{equation}
for $p$ small. For a large graph $p=1/2$ implies however
\begin{equation}\alpha\approx \frac{1}{2}\binom{n}{2}\gg
  n/2\end{equation}
That is, there is still a wide gap between these two values, a point
which needs further clarification.
\subsection{The Unfolded Epoch}
For $p$-values away from $p\approx 1$ we expect the
\tit{clique-graph}, $\mcal{C}$, to consist of a huge number of
cliques, $S_i$, each being surrounded by its \tit{local group}, that
is cliques having an overlap with $S_i$, while there is no longer an
overlap with the majority of cliques in the graph. In other words, in
this scenario the clique graph is much more unfolded and has at least
the potential to figure as a kind of \tit{proto space-time}. This
epoch shall now be analysed in more detail. 

From the general analysis we learned that the majority of cliques has
an order lying between $r_0/2$ and $r_0$, $r_0\approx\log(n)/\log(1/p)$. 
The expected number of $r$-cliques is $\langle Z_r
  \rangle=\binom{n}{r}\cdot(1-p^r)^{n-r}\cdot p^{\binom{r}{2}}$. We
  want the average order of cliques to be much larger than one, so
  that the internal structure of the lumps or physical points is still
  sufficiently complex but, on the other side, the order should be
 \tit{infinitesimal} compared to the number of nodes in the graph, that is
\begin{equation}1\ll r_0\lll n\end{equation}

In a first step we will now estimate the number of relevant cliques in
the network or graph for given bond-probability $p$ which implies an
$r_0$ of the above type (the numerical calculation follows below). We
approximate factorials by the following version of Stirling´s formula:
\begin{equation}x!\approx \sqrt{2\pi x}\cdot x^x\cdot
  e^{-x}\end{equation}
This yields
\begin{equation}\binom{n}{r}\approx \frac{n^n}{\sqrt{2\pi r}\cdot
    r^r\cdot (n-r)^{(n-r)}}\end{equation}
(the exponentials drop out)\\
As we have to deal with extremely large numbers it is useful to take
logarithms:
\begin{equation}\label{Stirling}  \log{\binom{n}{r}}\approx
  n\cdot\log(n)-(n-r)\log(n-r)\approx r\cdot\log(n)\end{equation}
where we have left out marginal contributions and approximated
$\log(n-r)$ by $\log(n)$. We hence have for the assumed range of
$(n,r)$-values:
\begin{equation}\binom{n}{r}\approx n^r\end{equation}
With this estimate we can now estimate $\langle Z_r\rangle$, the
expected number of $r$-cliques.
\begin{equation}\log(\langle Z_r\rangle)\approx
  r\cdot\log(n)+n\cdot\log(1-p^r)+r^2/2\cdot\log(p)\end{equation}
For the range of parameters we are interested in we can approximate
$\log(1-p^r)$ by $-\log(e)\cdot p^r$. Evaluating the latter expression
for e.g. the values of the list (\ref{tabelle1}), we see that
irrespective of the huge prefactor $n$ the second term in the above
equation becomes negligibly small (for e.g. $p=0.7$, $r=1000$ and
$n=10^{100}$ we have $n\cdot p^r \approx 10^{-54}$. Henceforth we will
therefore work with the formula
\begin{equation}\log(\langle Z_r\rangle)\approx
  r\cdot\log(n)+r^2/2\cdot\log(p)\end{equation}
To get an idea of the size of the numbers being involved we provide
the following list for the parameters $p=0.7$, $r_0=1333$, taken from
(\ref{tabelle1})
\begin{equation} \label{tabelle2}
\begin{array}{|c|c|c|c|c|c|c|}
r & 400 & 500 & 650 & 1200 & 1300 & 1400\\
\hline
\log(\langle Z_r\rangle) & 2.9\cdot 10^4 & 3.2\cdot 10^4 & 3.3\cdot
10^4 & 1.2\cdot 10^4 & 3\cdot 10^3 & 5\cdot 10^2
\end{array}
\end{equation}
\begin{ob}We see that over a large scale of $r$-values, ranging from,
  say, $100\le r\le 1200$, we have a $\langle Z_r\rangle$ of
  $O(10^{10^4})$. On the other side, we noted above that from
  general estimates (\cite{Random}) one can infer that the typical
  cliques are supposed to have $r$-values between $r_0/2$ and $r_0$.
  The above list shows that, at least for our possibly extreme values,
  this number is still appreciable below $r_0/2\approx 650$. On the
  other side, the upper critical value, i.e. $r_0\approx 1300$ seems
  to be more reliable.
\end{ob}  

In a next step we want to estimate the average number of cliques which
overlap with a given fixed clique, $S_0$ of a generic order between,
say, $r_0/2$ and $r_0$. This set of cliques we dubbed the \tit{local
  group} or \tit{infinitesimal neighborhood} of the respective lump
or physical point, $S_0$. This problem winds up to a calculation of
certain \tit{conditional probabilities}. As subset of graphs we take
those containing a fixed $r$-clique, $S_0$. The respective relative
probability (which is the same as the measure of the set of graphs
under discussion) is $(1-p^r)^{(n-r)}\cdot p^{\binom{r}{2}}$. In this
subclass of graphs we now consider another subset, consisting of the
graphs containing yet another $r'$-clique, having an overlap of order $l$
with the fixed given $S_0$.

The (conditional) probability that an $r$-set and an $r'$-set of nodes
span an $r$-clique, $S_0$, $r'$-clique, $S_1$, respectively, with common overlap of
order $l$ is
\begin{equation}\frac{p^{\binom{r}{2}}\cdot
    p^{\binom{r'}{2}-\binom{l}{2}}}{(1-p^r)^{n-r}\cdot
    p^{\binom{r}{2}}}\cdot P_{r',l}
\end{equation} 
with $P_{r',l}$ to be calculated in the following way. We have to
apply the same principle as in the calculation of individual cliques,
i.e. incorporate the fact that they are maximal members in the class
of $(r,r')$-simplices. On the other side, they now overlap and we have
to avoid a double counting of elementary events
(i.e. configurations). Maximality plus overlap is then encoded in the
following way:
\begin{equation}P_{r',l}=(1-p^{(r+r'-l)})^{(n-(r+r'-l))}\cdot
  (1-p^{(r-l)})^{(r'-l)}\cdot (1-p^{(r'-l)})^{(r-l)}\end{equation}
with $P_{r',l}$ describing the probability that there is no other node
in the graph under consideration which has all its links to nodes in
$S_0,S_1$ being occupied.

In this new probability space of graphs, containing a fixed
$r$-clique, $S_0$, there exist $\binom{r}{l}\cdot \binom{n-r}{r'-l}$
possibilities to place an $r'$-clique with overlap $l$. Hence we have
the result
\begin{conclusion} The expected number of $r'$-cliques, overlapping
  with a fixed given $r$-clique, $S_0$, is 
\begin{equation}\sum_{l=1}^{r-1}\langle N(S_0;r',l)\rangle
\end{equation}
with
\begin{equation}\langle N(S_0;r',l)\rangle=\frac{\binom{r}{l}\cdot
 \binom{n-r}{r'-l}\cdot
      p^{\binom{r'}{2}-\binom{l}{2}}}{(1-p^r)^{n-r}\cdot
      p^{\binom{r}{2}}}\cdot P_{r',l}
\end{equation}
\end{conclusion}

To estimate the cardinality, $\langle N_{loc.gr.}\rangle$, of the local group of $S_0$, one has first
to sum over all $r'$-values between, say, $r_0/2$ and $r_0$. We showed
in the above tabel,(\ref{tabelle2}), that the numbers are rather
uniform over a wide range of $r$-values so that it is sufficient to
estimate the cardinality for some generic value, $r'\approx r$ say, and
simply multiply this result with the width of the interval
(e.g. $10^3$ in the above table). In a first step we have to convince
ourselves that, as in the above numerical calculation, we can again
neglect the $P_{r',l}$-factors if we are only interested in estimates
of the kind \tit{order of}. Taking logarithms we see that all terms
arising from $\log(P_{r',l})$ are either very small or at most of
$O(1)$. The same holds for the term $(n-r)\cdot
\log(1-p^r)$. They can be neglected compared with the other terms and
we hence end up with
\begin{equation}\label{l}   \sum_{l=1}^{r-1}\langle N(S_0;r',l)\rangle\approx 
\sum_{l=1}^{r-1}\binom{r}{l}\cdot \binom{n-r}{r'-l}\cdot
p^{\binom{r'}{2}-\binom{l}{2}}
\end{equation}
On the other side, the expected number of $r'$-cliques \tit{not}
overlapping with $S_0$ is
\begin{equation}
\begin{split}
\langle N(S_0;r',l=0)\rangle &=\binom{n-r}{r'}\cdot
  p^{\binom{r'}{2}}\cdot(1-p^r)^{-(n-r)}\cdot P_{r',l=0}\\
& \approx \binom{n-r}{r'}\cdot p^{\binom{r'}{2}}
\end{split}
\end{equation}
which is for e.g. $r'\approx r$ of the same order as $\langle
Z_r\rangle$ itself (with $n\approx (n-r)$ for $n\ggg r$). In other
words, we observe already a certain degree of \tit{unfolding} on the
level of the \tit{clique graph} as compared to the underlying graph,
$G$. But nevertheless, as all the estimates are only of \tit{order-of}
type, we cannot use the latter estimate as a substitute for (\ref{l})
which can also be very large in principle.
\begin{bem}Previously we approximated $\log\binom{n}{r}$ in
  expressions like $\log(\binom{n}{r}\cdot p^{\binom{r}{2}})$ by
  $r\cdot\log(n)$, neglecting the term $-r\cdot\log(r)$. This suffices as long as $n\ggg r \gg 1$. But if
  $l\approx r'$ as in the above expression, we have to take a further
  term into account in the approximation of the Stirling formula (cf.
  (\ref{Stirling}).
\end{bem}
We have
\begin{multline}\log\langle N(S_0;r',l)\rangle\approx
  l\cdot(\log(r)-\log(l))+(r'-l)\cdot(\log(n)-\log(r'-l))\\
+1/2\cdot((r')^2-l^2)\cdot\log(p)\end{multline}
For $l\ll r'$ we can still neglect the correction term and get
\begin{equation}\log\langle
  N(S_0;r',l)\rangle\approx(r'-l)\cdot\log(n)+1/2\cdot
 (r')^2\cdot\log(p)\approx \log\langle Z_{r'}\rangle-l\cdot\log(n)\end{equation}
For $l\lessapprox r'$ (and $r,r'$ both generic, i.e. of roughly the
same order) we have on the other side
\begin{equation}\log\langle N(S_0;r',l)\rangle\approx
  l\cdot(\log(r)-\log(l))+(r'-l)\cdot(\log(n)-r'\cdot\log(p))
\end{equation}
which is much smaller than the preceding expression.
\begin{conclusion}[Typical Size of Local Group in Clique Graph
  ]\hfill\\
From the above we infer that the dominant contribution comes from
cliques having a relatively small overlap with the given lump,
$S_0$. Neglecting small factors and correction terms we have
approximately for generic $r,r'$:
\begin{equation}|\text{local group}|\approx \sum_{l\ll
    r'}|\text{number of cliques in $G$}|/n^l
\end{equation}
with the main contribution coming from $l=1$.
\end{conclusion}
That is, the size of the typical local group is roughly the $n$-th part of the
total number of cliques.
\begin{ob}From tables (\ref{tabelle1}) and (\ref{tabelle2}) we see
  that the local groups in the clique graph stemming from a typical
  random graph are still very large but much smaller than the
  original neighborhood of a node in the underlying graph, $G$.
\end{ob}
\begin{bem}If we assume that members of the local group should have an
  appreciable overlap (i.e. $l\gg 1$), the size would shrink
  drastically as can be seen from the above general formula.
\end{bem}

\subsection{Some Desirable Properties of (Proto) Space-Time}
What we have achieved so far is the following. We did not tackle the
full problem head-on, i.e. show that something like unfolded
(proto) space-time emerges under the evolution of a certain
\tit{critical} dynamical law from a chaotic initial phase. Instead of
that we replaced this (presumably) very complicated problem by a
slightly different one. A suitable statistical generic behavior
assumed, we described the emergence of a \tit{two-story structure}
within the framework of \tit{random graphs}. More precisely, we showed
that the \tit{wiring diagram} of the network, i.e. the underlying
(clock-time dependent) graph, passes generically through various
epochs with decreasing \tit{bond probability}, $p$, starting from an
\tit{embryonic epoch} with almost all \tit{cliques} overlapping for
$p$ in the infinitesimal neighborhood of one, and unfolding, for $p$
away from one, into a web of cliques (the \tit{clique graph}),
displaying a certain degree of \tit{near-/far-order}. In this epoch or
\tit{phase} each lump has a certain local neighborhood of other lumps
overlapping with it, these local groups now being significantly smaller
than the whole web.

We want, however, to stress the following point. The emergence of
something highly organized (from a dynamical and macroscopical point
of view) as macroscopic space-time \tit{cannot} expected to be a mere
probabilistic effect coming about as a byproduct without the injection
of a (possibly very peculiarly chosen) class of dynamical laws. To
learn more about the (un)naturalness of such a dynamical process, we
want to show in the remaining space what kind of results can already
be achieved in the pure random graph approach and compare them with
what is needed in order to have a fully developed space-time picture.

As to the underlying graph, $G$, we worked with numerical data like
\begin{equation}n=10^{100}\quad p=0.7\quad r_0=1333\end{equation}
The number of generic cliques was then approximately
\begin{equation}\langle Z_r\rangle\approx 10^{10^4}\end{equation}
The size of the typical local group of a clique was roughly
\begin{equation}\langle N_{loc.gr.}\rangle\approx \langle Z_r
  \rangle/n\end{equation}
which is, by the same token, the average \tit{vertex degree},
$\langle v_{cl}\rangle$, in the \tit{clique graph}, i.e. the average
number of bonds ending on a vertex.  

From this, one can estimate the bond probability, $p_{cl}$, in the
clique graph, with an existing bond meaning overlap.
\begin{equation}p_{cl}=(n-1)/\langle v_{cl}\rangle\approx \langle
  Z_r\rangle/\langle N_{loc.gr.}\rangle\approx n^{-1}\end{equation}
\begin{ob}With the numerical data we used above we have an edge
  probability in the clique graph
\begin{equation}p_{cl}\approx n^{-1}=10^{-100}\end{equation}
which is very small (compared to e.g. the $p=0.7$ of the underlying
graph).
\end{ob}
Another important question is whether the clique graph is generically
connected. The threshold we presented in \ref{connectedness} is
\begin{equation}p^*(n)=\log(n)/n\approx
  10^4/10^{10^4}=10^{-(10^4-4)}\ll p_{cl}\end{equation}
\begin{conclusion}For the numerical data we employed the web of lumps
  is \tit{almost surely} connected.
\end{conclusion} 

Take, on the other side, a typical class of graphs which are very
regular, admitting large or infinite distances between nodes and have
a pronounced near-/far-order, i.e. \tit{lattices}. Usually the vertex
degree, $v$, is both low and constant all over the lattice. For node
number, $n$, to infinity we hence have
\begin{equation}v\ll p^*(n)\cdot (n-1)=v^*\approx \log(n)\end{equation}
that is, a typical lattice graph is a \tit{connected} member of the
regime of random graphs which are generically \tit{disconnected}. This
shows that graphs which may be of particular interest for physics and
numerical approximation are, on the other side, not generic in a
probabilistic sense. In the following we show that, in fact, a too
large $\langle v\rangle$ or $p$ usually imply a very low
\tit{diameter} of the graph which is typically equal to two.  
\begin{defi}[Diameter] The diameter, diam(G), of a graph is defined as
  the greatest distance between two arbitrary nodes of the graph, i.e.
\begin{equation}diam(G)=\max_{i,j}d(n_i,n_j)\end{equation}
\end{defi}
Surprisingly, this diameter behaves rather uniformly over a wide range
of $p$-values in a random graph. In order to derive a quantitative
estimate we will proceed as follows. In a first step we calculate the
probability that an arbitrary pair of nodes, $(n_i,n_j)$, is neither
directly connected nor connected with each other via an intermediate
node, $x$, by a pair of bonds. For a fixed pair $n_i,n_j$ and $x$
running through the set of remaining nodes this probability is
(following the same line of reasoning as above) $(1-p)\cdot
(1-p^2)^{(n-2)}$.  There are $\binom{n}{2}$ such pairs. We denote by
$A$ the random variable (\tit{number of such pairs})
and have
\begin{equation}\langle A\rangle =\sum_{k\ge1}k\cdot
  pr(A=k)=\binom{n}{2}\cdot(1-p)(1-p^2)^{(n-2)}\end{equation}
On the other side
\begin{equation}\langle A\rangle\ge\sum_{k\ge1}pr(A=k)=pr(\exists\;
  \text{such a pair})=:pr(A)=1-pr(A=0)\end{equation}
with
\begin{equation}pr(A=0)=pr(diam(G)\le2)\end{equation}
Hence we have
\begin{ob}The probability that $diam(G)\leq 2$ is $1-pr(A)\ge
  1-\langle A\rangle$, which is of course only non-trivial for
  $\langle A\rangle<1$. 
\end{ob}
Remark: Note that there is exactly one graph having a diameter equal
to one, i.e. the simplex over the $n$ vertices. In the large-$n$-limit
this contribution becomes marginal.\vspace{0.5cm}

For $n$ large, more precisely, $n\to\infty$, one can now calculate the
$n$-dependent $p^*(n)$-threshold so that $diam(G)=2$ holds almost
surely for $p(n)>p^*(n)$ (cf. \cite{Random}). We content ourselves
with a simpler but nevertheless impressive result. With $p>0$ fixed and
$n\to\infty$ we infer from the above estimate
\begin{equation}\log(\langle A\rangle)\approx 2\log(n)-\log2+\log(1-p)+(n-2)\cdot\log(1-p^2)\end{equation}
which is for fixed $p$ and $n\to\infty$ dominated by the last term
which goes to $-\infty$.
\begin{conclusion}With $p>0$ fixed and $n\to\infty$ G(n,p) has
  diameter two almost shurely. Inserting now the numerical values for
  our \tit{clique graph} $\mcal{C}$ in the above formula, we see that
  the very small $p_{cl}$ is overcompensated by the huge $n_{cl}$ with the
  result that not only the underlying graph $G$ but also the derived
  clique graph $\mcal{C}$ has still diameter equal to two almost surely. 
\end{conclusion}

The above result is in fact not so bad, whereas it exhibits the limits
of the pure random graph approach. It rather shows that the unfolded
web of lumps we expect to emerge as a kind of (proto) space-time is
presumably the result of some dynamical fine tuning which, among other
things, accomplishes the possibility of larger distances between the
lumps. This fine tuning is expected to shrink the available \tit{phase
  space} so that the set of graphs we have to take into account is in
effect much smaller and more special. Note that on a continuum scale
the driving force which fuels the unfolding towards larger scales is a
combined effect of Einstein gravity and the (not completely
understood) behavior of the quantum vacuum. On our much more
primordial scale there will be a certain pendant in form of a peculiar
class of dynamical network laws, which may then lead, on e.g. the
scale of the web of lumps or a coarser scale, to an \tit{effective theory} which serves
as a proto form of macroscopic gravity (in the mentioned companion
paper we attempt to show this for quantum theory).

We want to remark that we recently came across a, as we think, deep
speculative remark by Penrose (\cite{Penrose65}) in which he ventured
the idea that, perhaps, some sort of \tit{(quantum) non-locality} is
already necessary on ordinary scales in order to understand the
formation of crystals and the like. In the case of quantum theory we
will argue that it just turns out to be such an effective theory, encoding
\tit{non-local} effects in a seemingly \tit{local} framework. We
expect the same kind of non-locality to be at work in the organisation
of our web of lumps.  

A network or graph characteristic which has the potential to describe
the large-scale organisation of our web of lumps is a concept we
introduced and analyzed in \cite{1}, i.e. the notion of
\tit{(intrinsic) graph dimension} or \tit{scaling dimension}. We do
not want to go into the details here as the topic poses difficult
questions of their own right. We only briefly discuss this concept and
relate it to the questions we are investigating in this paper. We
begin by providing two definitions of a possible graph dimension,
which turn out to be the same on generic configurations but may be different on
exceptional ones (cf. \cite{1}).
\begin{defi} With $U_k(x)$ describing the $k$-neigborhood of the node
  $x$, i.e. the nodes, $n_i$, with distance $d(x,n_i)\le k$, $\partial
  U_k(x)$ its boundary, i.e. the nodes having $d(x,n_i)=k$, we define
\begin{equation}D^1_k(x):= \ln|U_k(x)|/\ln(k)\end{equation}
\begin{equation}D^2_k(x):= \ln|\partial U_k(x)|/\ln(k)+1\end{equation}
and call
\begin{equation}\overline{D}:=\limsup D^{(1,2)}_k\quad , \quad
  \underline{D}:= \liminf D^{(1,2)}_k\end{equation}
the {\em upper, lower dimension} (for more details see \cite{1}). If
they coincide we call them the respective dimensions of the graph or network.
\end{defi}

We showed in \cite{1} that for, say, regular lattices, embedded in a
background space like $\R^n$, the above intrinsic dimensions coincide
with the usual euclidean dimensions of the embedding space (a
property well known to hold for \tit{fractal}-like dimensions as ours). On the
other side, we constructed examples (i.e. graphs) for practically
every (non-integer) value of $D^{(1,2)}$ in \cite{1}, showing that it
is in fact a useful generalisation of the ordinary continuous
dimension concept. Furthermore, we could show, among other things,
that $D$, provided it exists at all, is practically independent of the
special node $x$ but is rather a property of the whole graph.

From a physical, more specifically, \tit{dynamical}, point of view the
above dimensions exactly encode just the characteristics of the system
which really matter physically (e.g. in the context of phase
transitions). They count basically the number of \tit{new} partners
(nodes) which can be reached or influenced within \tit{clock-time}
$k,\;k\to\infty$. As the continuum space dimension of our univerese
seems to be rather special (presumably being even fine-tuned), it is
of central importance to understand how and why some definite (and
integer) dimension like e.g. three may emerge in our network. Our
approach makes it possible to analyze this great and longstanding
question in a dynamical and microscopic way. As this very intricate
subject poses a variety of subtle problems of their own, we refrain
from treating it here (due to limits of space).

We want to close the paper with emphasizing an aspect of our two story
structure of the network which will be of crucial importance for
fundamental physics in general and for understanding \tit{quantum
  theory} in particular. 
\begin{ob}[The Two-Story Concept]\hfill\\
\begin{enumerate}
\item Given a network or graph, $G$, of the above kind, we can
  construct its associated {\em clique graph}, $\mcal{C}_G$, and thus
  establish the two story concept, mentioned already in the
  introduction. We hence have two kinds of {\em distances} and {\em metric
    (causal) relations} in the network, the one defined by the
  original node distance in $G$, the other by the distance between
  lumps (defined by {\em overlap}) in $\mcal{C}_G$.
\item It is important that two lumps, $S_1,S_2$, which are some
  distance apart in $\mcal{C}_G$, may nevertheless be connected by a
  certain (possibly substantial) number of {\em interbonds} or short
  paths, extending from nodes in $S_1$ to nodes in $S_2$
 (see the construction of the cliques described in the preceding sections). 
\item That is, there may exist two types of information transport or correlation
  being exchanged in the network. A relatively coherent (and possibly
  {\em classical}) one, exchanged among the lumps, and a more
  stochastic and less coherent one (of possibly {\em quantum nature})
  between individual groups of nodes lying in the respective lumps and
  which can be almost {\em instantaneous}.
\end{enumerate}
\end{ob}
Remark: These phemomena will be employed and analyzed in the companion
paper about quantum theory. For some technical results concerning
degrees of interdependence of sets of nodes see the last section of \cite{Planck}

\end{document}